\documentclass[12pt]{article}
\usepackage{times}
\usepackage{geometry}
\usepackage[utf8]{inputenc}
\usepackage{enumitem,amssymb}
\usepackage{ragged2e}
\usepackage{parskip}
\usepackage[dvipsnames]{xcolor}
\usepackage{graphicx}
\newlist{thematic}{itemize}{8}
\setlist[thematic]{label=$\square$}
\usepackage{pifont}
\usepackage{xspace}
\usepackage[version=4]{mhchem}
\newcommand{\cmark}{\ding{51}}%
\newcommand{\done}{\rlap{$\square$}{\raisebox{2pt}{\large\hspace{1pt}\cmark}}\hspace{-2.5pt}}

\newcommand{\aap}{A\&A}
\newcommand{\araa}{ARAA}
\newcommand{\aj}{AJ}
\newcommand{\apj}{ApJ}
\newcommand{\apjl}{ApJL}
\newcommand{\mnras}{MNRAS}

\newcommand{\mum}{\,\ensuremath{\mathrm{\mu m}}\xspace}

\newcommand{\HI}{\ce{H\,\textsc{i}}\xspace}
\newcommand{\HII}{\ce{H\,\textsc{ii}}\xspace}
\newcommand{\CI}{\ce{[C\,\textsc{i}]}\xspace}
\newcommand{\CII}{\ce{[C\,\textsc{ii}]}\xspace}

\newcommand{\OI}{\ce{[O\,\textsc{i}]}\xspace}

\newcommand{\kms}{km s$^{-1}$\xspace}

\newcommand{\Htwo}{\ce{H2}\xspace}

\newcommand{\blue}{\textcolor{blue}}

\begin{document}
\raggedright
\huge
Astro2020 Science White Paper \linebreak

The Cycling of Matter from the Interstellar Medium to Stars and back \linebreak
\normalsize

\noindent \textbf{Thematic Areas:} \hspace*{60pt} $\square$ Planetary Systems \hspace*{10pt} \done Star and Planet Formation \hspace*{20pt}\linebreak
$\square$ Formation and Evolution of Compact Objects \hspace*{31pt} $\square$ Cosmology and Fundamental Physics \linebreak
  $\square$  Stars and Stellar Evolution \hspace*{1pt} \done Resolved Stellar Populations and their Environments \hspace*{40pt} \linebreak
  $\square$    Galaxy Evolution   \hspace*{45pt} $\square$             Multi-Messenger Astronomy and Astrophysics \hspace*{65pt} \linebreak
  
\textbf{Principal Author:}

Name:	Robert Simon
 \linebreak						
Institution:  Universität zu Köln
 \linebreak
Email: simonr@ph1.uni-koeln.de
 \linebreak
 
\textbf{Co-authors:} (names and institutions)
\linebreak
Nicola Schneider, Universität zu Köln (UzK);
Frank Bigiel, Universität Bonn; 
Volker Ossenkopf-Okada, UzK;
Yoko Okada, UzK;
Doug Johnstone, National Research Council Canada;
Peter Schilke, UzK;
Gordon Stacey, Cornell University;
Markus Röllig, UzK;
Álvaro Sánchez-Monge, UzK;
Daniel Seifried, UzK; 
Jürgen Stutzki, UzK;
Frank Bertoldi, Universität Bonn;
Christof Buchbender, UzK;
Michel Fich, University of Waterloo; 
Terry Herter, Cornell University;
Ronan Higgins, UzK;
Thomas Nikola, Cornell University

%

\textbf{Abstract:} \\

Understanding the matter cycle in the interstellar medium of galaxies from the assembly of clouds to
star formation and stellar feedback remains an important and exciting field in comtemporary
astrophysics.  Many open questions regarding cloud and structure formation, the role of turbulence,
and the relative importance of the various feedback processes can only be addressed with
observations of spectrally resolved lines.  We here stress the importance of two specific sets of
lines: the finestructure lines of atomic carbon as a tracer of the dark molecular gas and mid-J CO
lines as tracers of the warm, active molecular gas in regions of turbulence dissipation and
feedback.
The observations must cover a wide range of environments (i.e., physical conditions), which will be
achieved by large scale surveys of Galactic molecular clouds, the Galactic Center, the Magellanic
clouds, and nearby galaxies. To date, such surveys are completely missing and thus constitute an
important science opportunity for the next decade and beyond.
For the successful interpretation of the observations, it will be essential to combine them with
results from (chemical) modelling and simulations of the interstellar medium.






\pagebreak
\textbf{\large 1. Scientific context} 

The science case outlined in this white paper is built around the question:
\textit{How do the processes involved in star formation and stellar feedback
  shape the interstellar medium in galaxies?}


{\bf 1.1 The cyle of matter} \\

The cycling of matter in the interstellar medium (ISM) begins with the accretion of gas onto
galactic disks and cooling to form a neutral phase. It progresses with the formation of molecular
clouds out of the diffuse, atomic gas and the formation of denser sub-structures such as filaments
and cores which ultimately form stars and clusters. These stars in turn interact with the ambient
ISM via feedback (radiation, mechanical, supernovae), shaping the ISM properties and chemically
enriching galaxies over cosmic time.

{\bf 1.2 Molecular cloud and star formation} \\

How in detail molecular clouds, dense structures within them, and stars are forming remains disputed
and is a highly active field in observational and theoretical astrophysics.  We promote here a
scenario supported by numerical simulations (e.g., V\'azquez-Semadeni et al. 2006,
Heitsch \& Hartmann 2008)
in which molecular clouds assemble fast from converging \HI flows in
the warm neutral medium.
Turbulence plays a key role because it supports clouds at large scales but also creates a highly
inhomogeneous molecular cloud structure that is characterized by large density contrasts (Mac Low \&
Klessen 2004).  Most of these density fluctuations are transient, but where turbulent energy is
dissipated and shocks are present (Neufeld \& Dalgarno 1989, Godard et al. 2019), dense structures
in the form of filaments can emerge. Fragmentation on small scales then leads to the formation of
individual stars and star clusters.
The observed broad molecular lines (Goldsmith et al. 2008) and the complex structures seen in the
dust continuum support the omnipresence of supersonic turbulence in whole molecular clouds and even
larger scales.

{\bf 1.3 Injection of turbulence into the ISM} \\

Turbulence in molecular clouds must be constantly replenished or it would decay on time scales of
the order of the crossing time. In spite of the ubiquity of interstellar turbulence, the main
driving mechanisms are still poorly understood.  The observed network of filaments can be
qualitatively explained by numerical simulations of magneto-hydrodynamic turbulence (e.g., Padoan \&
Nordlund 2011, Federrath \& Klessen 2013) in gravitationally infalling gas.  Supersonic turbulence
has been attributed to a number of factors, including magnetic fields, protostellar outflows,
\HII\ regions, supernovae, and on-going mass accretion.
Observational tests have shown that the first four mechanisms are relatively inefficient. In their
theoretical study of the evolution of molecular clouds, Goldbaum et al. (2011) proposed that the
accretion of new material from the surrounding environment can drive the observed turbulent motions,
but this has not been thoroughly investigated observationally yet.
This study is also consistent with observations that suggest the global filamentary structure of molecular
clouds is created by large scale colliding flows of atomic material at earlier times (Walch et
al. 2015, Seifried et al. 2017).

\textbf{1.4 Heating and cooling in the ISM} \\

The thermodynamics of the gas
depends on the balance between heating and cooling processes, probably moderated by magnetic fields,
on large and small scales.
%
%
The most important processes are provided by stellar feedback (radiation, wind), low- and
high-velocity shocks, and cosmic-rays/X-rays. Cooling happens predominantly via dust and line
emission at (sub)mm, far-, and mid-infrared (IR) wavelengths so that studying this wavelength range
is the key to understand the relevant processes related to Galactic and extragalactic cloud and star
formation.

{\sl Understanding the matter cycle in the interstellar medium of galaxies from the assembly of
  clouds to star formation and stellar feedback remains an important and exciting field in
  comtemporary astrophysics. To achieve significant progress is truly a multi-scale and
  multi-physics problem that needs to be addressed in the coming decade from the observational and
  theoretical side.}

%




\textbf{\large 2. State of the art}

To study the cloud and star forming ISM in the Milky Way and external galaxies over large spatial
scales and different environmental conditions requires sensitive, high spatial resolution
observations in dedicated tracers.
The past decades have seen important advancements in multi-wavelength observations of the ISM driven
by the access to better (higher) sites (including air- and space-borne observatories) and
improvements in technology (higher sensitivity, multiplexing).  The routine operation of the Atacama
Large Millimeter Array (ALMA) is clearly a highlight in high-angular resolution submm
astronomy. ALMA covers a broad science case,
including observations of far-infrared (FIR)
cooling lines in high-redshifted galaxies. ALMA, however, can not efficiently survey large (many square degree) regions of the sky.

{\bf Dust continuum surveys} of large parts of the Milky Way, ranging from the infrared to the
submm, have become available thanks to satellite missions such as Spitzer or Herschel (see Fig.~1)
and ground-based telescopes (e.g., IRAM 30~m, JCMT, CSO, APEX). These data sets are invaluable to
determine the dust temperature and column density structure of the ISM.
However, they do not provide information on the dynamical state of the gas, the chemical state of
the gas, or even the phase of the gas. To obtain this, we need spectroscopic measurements of a suite
of molecules, atoms, and ions. \\
{\bf Wide field spectral line mapping} in the Milky Way is so far limited to atomic hydrogen (e.g.,
VLA and ATCA) and low energetic CO lines (starting with the pioneering Columbia survey (Dame et
al. 2001) and taken to higher angular resolution with, e.g., FCRAO, Mopra, JCMT, Nobeyama,
APEX). They trace the diffuse, neutral atomic and the cold and moderately dense molecular gas,
respectively. \\
The past 2 decades have also seen enormous progress in high-resolution, cloud-scale
($\sim$ 100 pc and below) dust continuum and CO line observations of external galaxies (e.g., IRAM
30~m/HERACLES (Leroy et al. 2009), ALMA/PHANGS (Sun et al. 2018)), including the Magellanic
Clouds. These studies allowed to assess the importance of environmental factors such as metallicity
for star formation.

For a deeper understanding of the processes forming molecular clouds and stars, and how this links
to galaxy evolution, it is essential to {\bf combine observations with modelling and simulations}.
Over the last 30 years, the main focus was on modelling Photon Dominated Regions (PDRs). Starting
with plane-parallel geometries (Hollenbach \& Tielens 1997, Sternberg \& Dalgarno 1989), PDR models
now incorporate more complex geometries (Röllig et al. 2006, Andree-Labsch et al. 2017) and
turbulence (Wolfire et al. 2010, Bialy et al. 2017). In parallel, considerable progress was made in
galaxy-wide modelling (Dobbs et al. 2008, Inoue \& Inutsuka 2012, Smith et al. 2014). However, only
coupling the chemistry with (magneto)-hydrodynamics, including self-gravity, variations in
metallicity, UV radiation and cosmic-rays has the potential for realistic modeling of the ISM on a
large scale. Because of increased computing power and laboratory work in molecular chemistry, these
models have become available (Walch et al. 2015, Kim \& Ostriker 2017).  With the inclusion of
radiative transfer, synthetic spectral line maps (Seifried et al. 2017, Franeck et al. 2018, Clark
et al. 2018) of the most important cooling lines of \CII, \CI, and CO now can directly be compared
to observations. To overcome the still existing discrepancies between observations and predictions
from simulations requires an intense synergy between modellers and observers.

\begin{figure}[t]
\includegraphics[width=\textwidth]{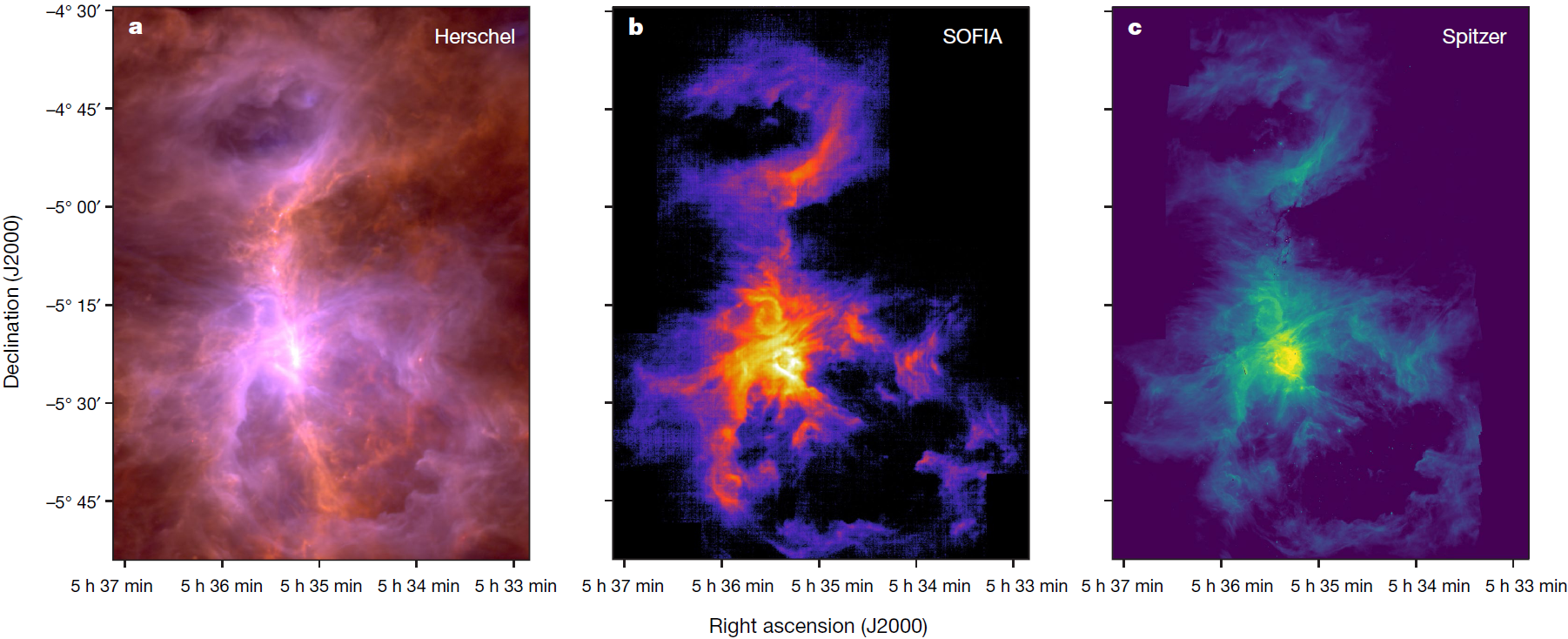}
\caption{The Orion A molecular cloud in the dust continuum (left: FIR/submm, right: mid-IR} and \CII spectral line emission (middle). From Pabst et al. 2019.
\end{figure}

\textbf{\large 3. Future opportunities}

The assembly of mass to form clouds, dense structures, and stars, and the impact of stellar feedback
are highly dynamic processes that require spectrally resolved observations of atomic and molecular
lines.  Wide-field spectral line surveys, however, are still hampered by either low angular
resolution, limited sky coverage or sensitivity, and, as we will specifically argue below, lack of
important atomic and molecular tracers.

Novel surveys employing large format (several 100 pixels), high sensitivity heterodyne arrays open
completely new opportunities: {\bf large scale, unbiased surveys} covering hundreds of square
degrees, possibly even the whole Galactic Plane and nearby galaxies. Such observations will allow
unique studies not possible in the past (e.g., of arm versus interarm gas, the effects of varying
cosmic ray fluxes, etc.) and a wide range of physical conditions (e.g., density, temperature,
pressure, radiation field, cosmic rays, metallicity, star formation activity/rate). And they will
allow to explore the unexpected.

In this context, {\bf molecular clouds of the Milky Way} provide the opportunity to study cloud
assembly and the highest possible spatial resolution to trace the details of structure formation,
turbulence dissipation, the formation of stars out of the densest gas, and the feedback of stars
into the ISM.  The {\bf Galactic Center and its Central Molecular Zone} (CMZ) serves as template for other
galactic nuclei. The overall conditions (temperature, density, pressure, radiation field) in the CMZ
are quite similar to other galactic nuclei, although the star formation rate in the CMZ at the
moment is rather low.
Observations of {\bf nearby galaxies} provide important context to Milky Way studies
as they expand the range of astrophysical environments (e.g., star formation activity, metallicity,
dynamics) and provide insight into the role of host galaxy properties related to molecular cloud and
star formation. High angular and spectral resolution is important to resolve processes in the Milky
Way and nearby galaxies spatially and dynamically to understand global properties of distant
galaxies which we will not be able to resolve even with the best instruments.

{\bf 3.1 The formation of molecular clouds} \\

It is uncertain how molecular clouds gain their mass as the relative role of the inflow of material
from high Galactic latitudes onto the gravitational potential of the disk and compression of gas in
spiral density waves or expanding shells of atomic hydrogen produced by supernovae is unknown.

Low energy CO surveys reveal the distribution of the cold, moderately dense \Htwo gas, but chemical
models predict that the formation of CO lags behind that of \Htwo and consequently, a different
tracer is required to account for this CO-dark \Htwo gas.  Simulations (Clark et al. 2018) show that
this gas emits in the fine structure lines of atomic and ionized carbon \CI and \CII.  While \CII is
emitted from all phases of the interstellar medium, \CI is specific to the CO-dark molecular gas as
it traces only the outer layers or early stages of molecular clouds. Franeck et al. (2018) showed
that \CII is mainly emitted from the atomic, not the CO-dark molecular gas.  Only when combining CO
observations with those of \CI as a tracer of the dark gas, we obtain a full picture of the
molecular material, in particular including the cloud mass accretion via low density \Htwo.

Compared to the 21 cm line of atomic hydrogen, those lines have the advantage that they are narrow
enough to resolve individual gas streams separated in velocity space. In contrast to \HI, \CI shows
much less line of sight confusion.  To measure the kinetic energy injected into the turbulent eddies
through accretion, a velocity resolution better than 1 \kms is required.  Due to the random nature
of turbulence, a conclusive comparison between models and observations is only possible with a
statistically significant sample of large-scale observations, ideally of the full Galactic Plane in
\CI and CO lines.

\textbf{To date, spectrally resolved \CI and mid-J CO surveys of large areas of the Galactic
  Plane and external galaxies are completely missing!} \\

%

Such surveys and dedicated smaller mapping projects, possibly at even higher observing frequency,
will become possible within a few years with the CCAT-prime observatory (Stacey et al. 2018) and are
planned for later ground based observatories (ATLAST at higher angular resolution) and
space missions (e.g., OST).  CCAT-prime is a 6~m aperture, very high surface accuracy, very large
field of view (2 degrees in diameter at 809 GHz) telescope at the superb (5600 m elevation
overlooking the ALMA plateau) Cerro Chajnantor site. CCAT-prime, together with a large format
heterodyne receiver such as CHAI (Stacey et al. 2018) can successfully complete the science
described here at better than 30" angular resolution in the next decade.

{\bf 3.2 Star formation and feedback} \\

While active star forming regions predominantly cool via the far- to mid-IR dust continuum, line
cooling is dominated by the fine structure lines of \CII and \OI at 158 and 63 \mum. These two lines
can only be observed from air- or space-borne platforms and the Stratospheric Observatory for
Far-Infrared Astronomy (SOFIA) is currently the only available facility. Large scale observations in
the \CII line are becoming available for whole molecular clouds and even galaxies (Pabst et al. 2019
for the Orion A cloud (Fig.~1) and Pineda et al. 2018 for M51). 

A significant fraction of the cooling is nevertheless in mid- to high-J CO lines, some of which are
accessible from the ground.
Observations of \CII, \CI and mid-J CO lines thus have the power to discriminate quiescent
(typically seen in low-J CO and \CI lines) from active gas in stellar feedback regions, which
enables the detailed study of PDRs, shocks (to be complemented by \OI
observations), outflows (in particular pc scale outflows that interact mostly with the neutral
atomic and ionized medium), collapse, and signatures of low-velocity shocks due to turbulence
dissipation, the ``smoking gun'' for the formation of dense structures.

In addition, sampling the Spectral Line Energy Distribution (SLED) towards higher J-lines (i.e.,
higher energies) is important to get an accurate handle on excitation and
cloud masses. Similarly, it is important to also observe the higher energetic \CI (2-1) line in
order to unambiguously determine the total amount of carbon.  Both provide valuable input for
chemical models of molecular clouds and PDRs, which are indispensible for the interpretation of the
observations.

{\bf 3.3 The galactic context} \\

The diverse population of nearby galaxies (Milky Way analogues, massive disk galaxies,
low-metallicity dwarfs, interacting systems) extend significantly the phase space of conditions
found in the ISM of the Milky Way.

Key sub-mm science cases for extragalactic observations in the future include resolved, full-disk,
kpc-scale observations of \CI and mid-J CO lines of a diverse sample of nearby
galaxies. Observations covering the low-J CO lines are available for many such targets (e.g., from
the IRAM 30~m, JCMT, ALMA, SMA), as are ancillary muli-wavelength data (e.g., dust and star
formation rate tracers) and increasingly also wide-area \CII\ mapping (e.g., from Herschel KINGFISH,
Smith et al. 2017, or SOFIA, Pineda et al. 2018, Bigiel et al. in prep.). \CI\ observations are rare
in nearby galaxies; a comprehensive \CI\ survey in combination with existing CO and \CII\ data thus
yields a complete carbon census resolved across galaxy disks and will provide the most accurate
calibration of these extragalactic molecular gas tracers to date. This is of particular importance
in light of \CI\ being routinely detected at high redshift (e.g., Wei\ss et al. 2005, Popping et
al. 2017).  \CI\ has furthermore been proposed as an effective dense gas tracer (and a viable
alternative to popular, extragalactic high critical-density lines like HCN or HCO$^+$, e.g.,
Papadopoulos \& Geach 2012).

Other important applications are the role of \CI as an optically thin molecular gas tracer (e.g.,
Papadopoulos \& Greve 2004), in particular in environments where CO fails as an effective
\Htwo\ tracer, e.g. at low metallicity. As the relative emissivities of \CI, \CII\ and CO in
molecular clouds changes with metallicity, wide area \CI surveys of the LMC and SMC will provide a
first systematic study of this ratio at low metallicity.

These science cases are also immediately related to observations at high redshift, where both
\CI\ and mid-J CO lines are routinely observed due to them being shifted into favorable bands with
sub(mm) interferometers like ALMA or NOEMA. A careful calibration in diverse environments in the
local universe is thus the key link between detailed, sub-cloud scale work in the Milky Way and
distant galaxies.



%
%

\pagebreak

\textbf{References} \\
Andree-Labsch, S., Ossenkopf-Okada, V., \& R{\"o}llig, M.\ 2017, \aap, 598, A2 \\

Bialy, S., Burkhart, B., \& Sternberg, A.\ 2017, \apj, 843, 92 \\

Clark, P.~C., Glover, S.~C.~O., Ragan, S.~E., \& Duarte-Cabral, A.\ 2018, arXiv:1809.00489 \\

Dame, T.~M., Hartmann, D., \& Thaddeus, P.\ 2001, \apj, 547, 792 \\
Dobbs, C.~L., Glover, S.~C.~O., Clark, P.~C., \& Klessen, R.~S.\ 2008, \mnras, 389, 1097 \\


Federrath, C., \& Klessen, R.~S.\ 2013, \apj, 763, 51 \\
Franeck, A., Walch, S., Seifried, D., et al.\ 2018, \mnras, 481, 4277 \\

Godard, B., Pineau des For{\^e}ts, G., Lesaffre, P., et al.\ 2019, \aap, 622, A100 \\
Goldbaum, N.~J., Krumholz, M.~R., Matzner, C.~D., \& McKee, C.~F.\ 2011, \apj, 738, 101 \\
Goldsmith, P.~F., Heyer, M., Narayanan, G., et al.\ 2008, \apj, 680, 428 \\

Heitsch, F., \& Hartmann, L.\ 2008, \apj, 689, 290 \\
Hollenbach, D.~J., \& Tielens, A.~G.~G.~M.\ 1997, \araa, 35, 179 \\

Inoue, T., \& Inutsuka, S.-i.\ 2012, \apj, 759, 35 \\
Kennicutt, R.~C., Calzetti, D., Aniano, G., et al.\ 2011, PASP, 123, 1347 \\
Kim, C.-G., \& Ostriker, E.~C.\ 2017, \apj, 846, 133 \\

Leroy, A.~K., Walter, F., Bigiel, F., et al.\ 2009, \aj, 137, 4670 \\

Mac Low, M.-M., \& Klessen, R.~S.\ 2004, Reviews of Modern Physics, 76, 125 \\

Neufeld, D.~A., \& Dalgarno, A.\ 1989, \apj, 344, 251 \\

Pabst, C., Higgins, R., Goicoechea, J.~R., et al.\ 2019, Nature, 565, 618 \\
Padoan, P., \& Nordlund, {\AA}.\ 2011, \apj, 730, 40 \\
Papadopoulos, P.~P., \& Geach, J.~E.\ 2012, \apj, 757, 157 \\
Papadopoulos, P.~P., \& Greve, T.~R.\ 2004, \apjl, 615, L29 \\
Pineda, J.~L., Fischer, C., Kapala, M., et al.\ 2018, \apjl, 869, L30 \\
Popping, G., Decarli, R., Man, A.~W.~S., et al.\ 2017, \aap, 602, A11 \\

Röllig, M., Abel, N.~P., Bell, T., et al.\ 2007, \aap, 467, 187 \\

Seifried, D., Walch, S., Girichidis, P., et al.\ 2017, \mnras, 472, 4797 \\
Smith, R.~J., Glover, S.~C.~O., \& Klessen, R.~S.\ 2014, \mnras, 445, 2900 \\
Smith, J.~D.~T., Croxall, K., Draine, B., et al.\ 2017, \apj, 834, 5 \\
Stacey, G.~J., Aravena, M., Basu, K., et al.\ 2018, Ground-based and Airborne Telescopes VII, 10700, 107001M \\
Sternberg, A., \& Dalgarno, A.\ 1989, \apj, 338, 197 \\
Sun, J., Leroy, A.~K., Schruba, A., et al.\ 2018, \apj, 860, 172 \\


V{\'a}zquez-Semadeni, E., Ryu, D., Passot, T., Gonz{\'a}lez, R.~F., \& Gazol, A.\ 2006, \apj, 643, 245 \\

Walch, S., Girichidis, P., Naab, T., et al.\ 2015, \mnras, 454, 238 \\
Wei{\ss}, A., Downes, D., Henkel, C., \& Walter, F.\ 2005, \aap, 429, L25 \\
Wolfire, M.~G., Hollenbach, D., \& McKee, C.~F.\ 2010, \apj, 716, 1191 


\end{document}